\documentclass[]{interact}

\usepackage{epstopdf}% To incorporate .eps illustrations using PDFLaTeX, etc.
\usepackage[caption=false]{subfig}% Support for small, `sub' figures and tables

\usepackage[longnamesfirst,sort]{natbib}% Citation support using natbib.sty
\bibpunct[, ]{(}{)}{;}{a}{,}{,}% Citation support using natbib.sty
\renewcommand\bibfont{\fontsize{10}{12}\selectfont}% To set the list of references in 10 point font using natbib.sty

\theoremstyle{plain}% Theorem-like structures provided by amsthm.sty

\theoremstyle{definition}

\theoremstyle{remark}

\usepackage{amsmath,amssymb,amsfonts}
\usepackage{algorithmic}
\usepackage{graphicx}
\usepackage{textcomp}

\usepackage{algorithm}
\usepackage{algorithmic}

  \usepackage{rotating}
        \usepackage{multirow}
        \usepackage{lscape}

\usepackage{xcolor}

\begin{document}

\articletype{}% Specify the article type or omit as appropriate

\title{Natural Language Processing via LDA Topic Model in Recommendation Systems}

\author{
\name{Hamed Jelodar \thanks{H. Jelodar. Email: Jelodar@njust.edu.cn}, Yongli Wang ,  Mahdi Rabbani, SeyedValyAllah Ayobi}
\affil{\textsuperscript{ Department of Computer Scicne Nanjing University of Scince and Technology, Nanjing, China}  }
}

\maketitle

\begin{abstract}
Today, Internet is one of the widest available media worldwide. Recommendation systems are increasingly being used in various applications such as movie recommendation, mobile recommendation, article recommendation and etc. Collaborative Filtering (CF) and Content-Based (CB) are Well-known techniques for building recommendation systems. Topic modeling based on LDA, is a powerful technique for semantic mining and perform topic extraction. In the past few years, many articles have been published based on LDA technique for building recommendation systems. In this paper, we present taxonomy of recommendation systems and applications based on LDA. In addition, we utilize LDA and Gibbs sampling algorithms to evaluate ISWC and WWW conference publications in computer science. Our study suggest that the recommendation systems based on LDA could be effective in building smart recommendation system in online communities.

\end{abstract}


\section{Introduction}
\label{sec1}
Today, Internet is one of the widest available media worldwide. It has essentially become a huge hit of data that has the potential to serve many information centric applications in our life. Recommendation system takes an essential part of many internet services and online applications, including applications like social-networking and recommendation of products (films, music, articles,..i.e.). Recommendation techniques have been used by the most known companies such as Amazon, Netflix and eBay to recommend releated items or products by estimating the probable preferences of customers. These techniques are profitable to both service provider and user. According to pervious works, two popular approaches for building recommendation systems can be categorized as content-based (CB), collaborative filtering (CF).
\begin{itemize}

\item Content-based (CB) recommending is adopted for recommendation systems model widely, which takes advantage of the property of items to create features and characteristics to coordinate user profiles. It can be relied on the properties of the items that each user likes to discover what else the user may like. One major issue of CB filtering method is that the recommendation system is required to gain an understanding of user preferences for some sorts of items and deploy these for other sorts of items.

\item Nevertheless CF has two widely known problems which are sparsity and cold start (CS). In the rating matrix, The percentage of elements which get values is small. Even it is possible that CF considers only a few rating for popular items. For instance, upon a considerable Netflix rating dataset which is provided for Netflix Prize competition, there are about 100 milion ratings for about 18,000 movies that are given by 480,000 users. The percentage of rating matrix elements which are received ratings is 1. With a sparse ranking matrix it is very challenging topic to make an effective recommendation, depending on estimation of the relationship between items and users. CS problem is another widely known issue for CF approach, which can occur on new users or items. In terms of achieving an effective recommendation, CF approach requires either ratings on an item or a large number of ratings from a user.

\end{itemize}

Recently, researchers proposed various methods based on probabilistic topic modeling methods \cite{1}. LDA is a generative probabilistic model broadly used in the information retrieval field. Researchers have used topic modeling methods based on LDA for building recommendation systems in various subjects, including app recommendation \cite{53,54,55,56,57}, event recommendation\cite{58,59,60,61,62,63,64,65,66,67}, hashtag recommendation \cite{28,29,30,31,32,33,34,35,36,37,38,39,40,41,42,43,44,45,46,47,48}, social networks and media \cite{22, 50, 58, 61, 77,78,79}. In this paper, We present a taxonomy of recommendation systems applications based on topic modeling (LDA) of the recent research  and evaluate ISWC and WWW publications in computer science between 2013 to 2017 from DBLP dataset.

\section{Natural language Processing and LDA topic model}
Topic models are a powerful and practical tool for analyzing huge text documents in Natural language processing. Topic models can automatically cluster words into topics and discover relationship between documents from a dataset. For example; we can assume a three-topic model of a News Dataset, including "sport", "money" and “politic”. The most common words in the sport topic (Topic 1) might be “gym”, “football”, and “tennis” and in addition, for politic topic (Topic 2) might be “senator”, “president”, and “election”; while the money topic (Topic 3) can be made up of words such as “dollar”, “currency”, and “euro”. Figure 1; show a simple example for understanding a topic discovery from group of words. LDA is a popular technique to semantic analysis in topic modeling and text mining. LDA can apply in a diversity of text-information to evaluate topic trends over time and analyze large numbers of documents.\\

\begin{figure*}
 \centering \includegraphics[scale=.5]{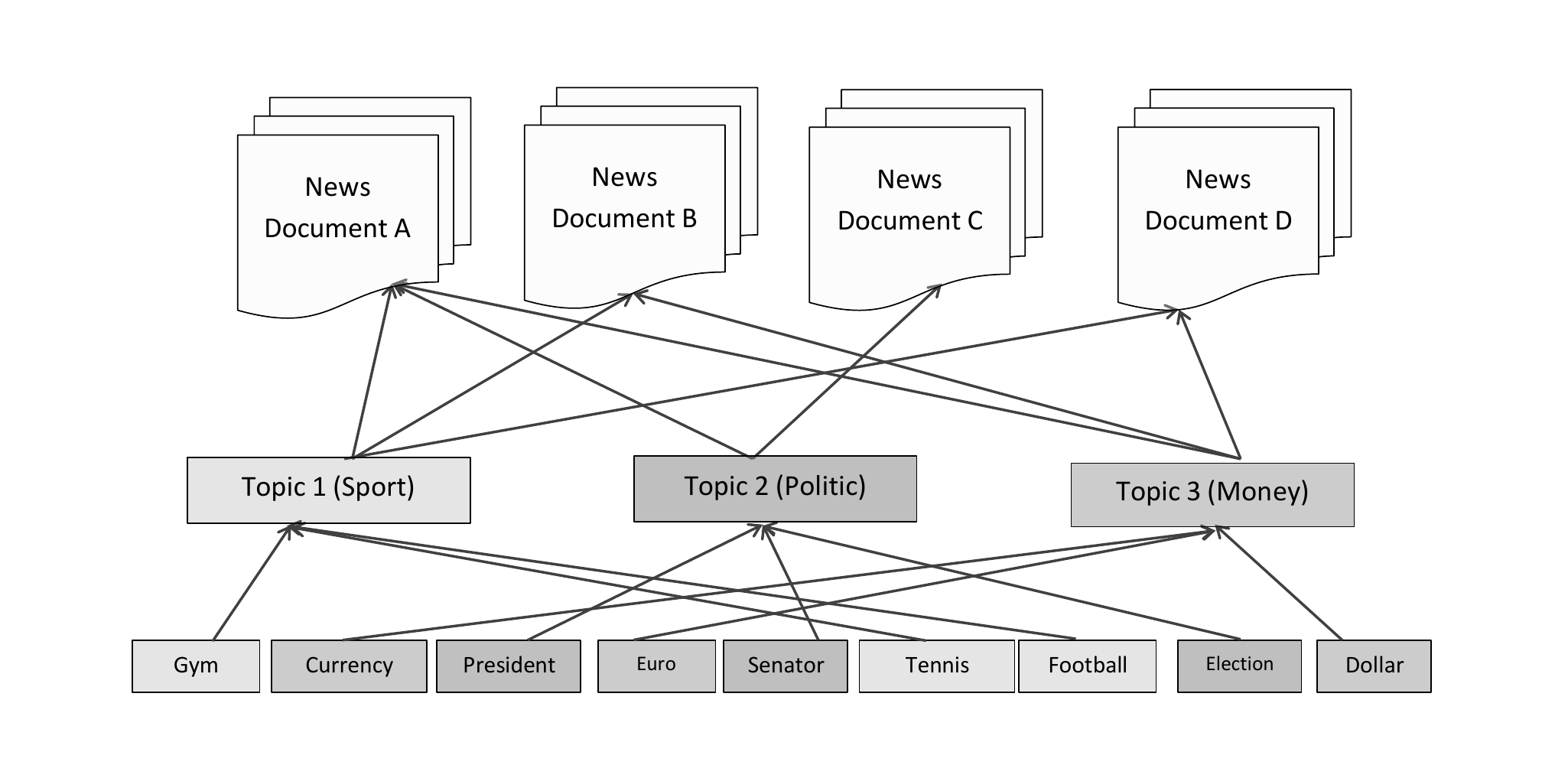}
\caption{As a simple example, showed the distribution of words on topics as well as the distribution of topics on documents as a simple example in topic modeling.}
\label{fig:1}     % Give a unique label
\end{figure*}

 In-process detail for LDA, defined a corpus (text) as $D={d_{1},d_{2},d_{3},....,d_{M}}$ where M is number of text documents and $d_{M}$ is a number of text documents in the corpus. A document is a series of N words denoted by $W={w_{1},w_{2},w_{3},....,w_{N}}$  , where $w_{N}$ is the $Nth$ word in the sequence of text document. In addition, z is a latent variable representing the hidden topic associated with each showed word.  The generative procedure for LDA, formally defined as:\\

\begin{itemize}
  \item{For topic index  $k\in\left\{1,2,...,k\right\}$}
    \subitem{i. Selected a word distribution  $\beta_{k}~Dir\left(\eta\right)$}
  \item{For text document $ D\in\left\{1,2,...,d\right\}$}
      \subitem{i.Selected a topic distribution  $\_{k}~Dir\left(\eta\right)$}
      \subitem{ii.For $n\in\left\{1,....,N_{d}\right\}$ word}
         \subsubitem{a.Selected a topic assignment $z_{dn}~Mult\left(\theta_{d}\right)$}
         \subsubitem{b.Selected a word $w_{dn}~Mult\left(\beta_{z_{d,n}}\right)$}

\end{itemize}
Mult() is a multinomial distribution, and $Dir()$ is a Dirichlet distribution which is a prior distribution of Mult(),$\alpha$ and $\theta$ are hyperparameters.\\

\subsection{Gibbs sampling and Learning LDA}
As previously mentioned, Topic modeling can find a collection of distributions over words for each topic and the relationship of topics with each document. To perform approximate inference and learning LDA, there are many inference methods for LDA topic model such as Gibbs sampling, collapsed Variational Bayes, Expectation Maximization. Gibbs sampling is a popular technique because of its simplicity and low latency. However, for large numbers of topics, Gibbs sampling can become unwieldy. In this paper, we use Gibbs Sampling in our experiment in section 5.

\section{Topic model based on recommendation systems: Recent research} \label{sec:2}
In this section, we considered six recommendation systems based on LDA which includes:
scientific paper recommendation \cite{2,3,4,5,6,7,8,9},
 music and video recommendation \cite{10,11,12,13,14,15,16,17,18,19,20},
 location recommendation\cite{21,22,23,24,25,26,27},
  travel and tour recommendation\cite{49,50,51,52},
    app recommendation\cite{53,54,55,56,57},
      friend recommendation \cite{79,69,80}, as shown in Figure 2.

\begin{figure*}
 \centering \includegraphics[scale=.5]{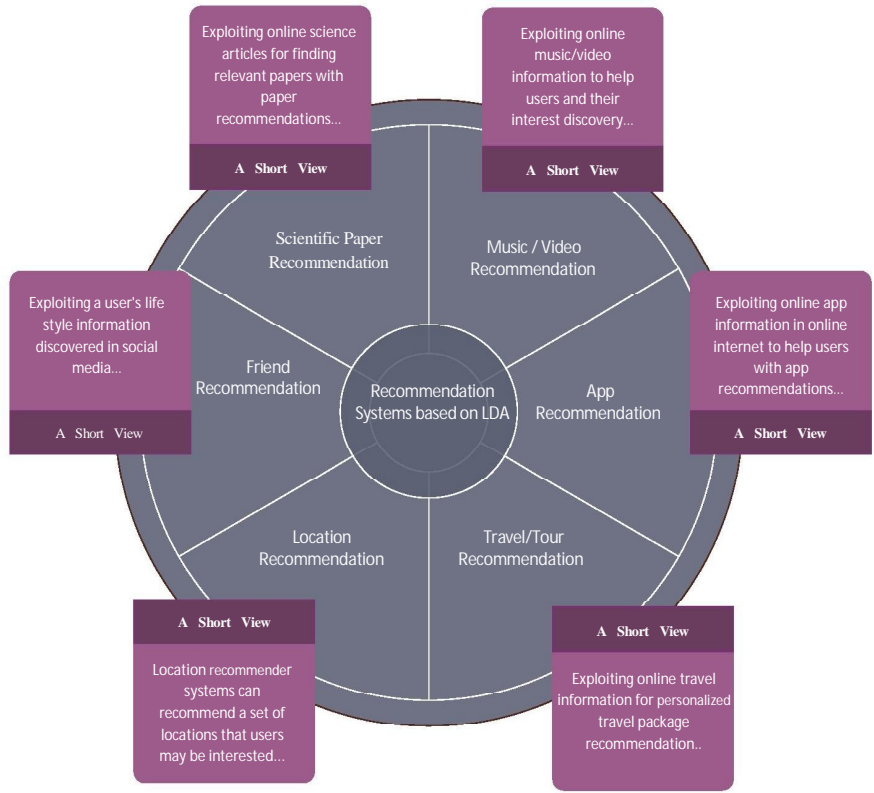}
\caption{A taxonomy of recommendation systems application based on LDA in various subject of
the recent research.}
\label{fig:1}     % Give a unique label
\end{figure*}

\subsection{Topic model based on scientific paper recommendation}

In recent years a considerable amount of research has addressed the task of defining models and systems for scientific papers recommendation; this trend has emerged as a natural consequence of the increasing growth of the number of scientific publications. For example, Youn and et al, proposed an approach to scientific articles'recommendation of user's interests based on a topic modeling framework. The authors, used a LDA model in order to extract the topics of the followees'tweets (followed Twitterers) and the paper titles\cite{9}. They apply the Twitter-LDA algorithm simultaneously on the followees' tweets and the paper titles with the number of topics set to 200, they utilized the intersection of topics found in both paper titles and followees's tweets. Each followee of a user is ranked as follows:

\begin{equation} \label{GrindEQ__5_}
{Rank}_{f\textrm{f}\textrm{o}llowee}=\left(\sum_{t\in {Topics}_p}{\frac{n\left(t,T_f\right)}{\left|T_f\right|}}\right)*|{Topics}_p\bigcap {Topics}_f|
\end{equation}

where $T_f$denotes all tweets by a followee,

${Topics}_p,\ $denotes the set of topics defining the titles of scientific
articles,

${Topics}_f,\ $denotes the set of topics defining the tweets of a followee
and\textbf{ }$n\left(t,T_f\right)\ $the number of times a particular topic
\textbf{`t'} from within occurs among the tweets of a followee. Based on the
ranking scores of all followees of a particular user,and obtained top-k researchers
followed by a target user. For evaluation approach, considered DBLP database as a
large academic bibliographic network.\\

% For tables use - t1
\begin{table}

\centering
\caption{Impressive works LDA-based on paper recommendation}
\label{tab:1}       % Give a unique label

\resizebox{12cm}{!} {

\begin{tabular}{lllll}
\hline\noalign{\smallskip}
Study & Year & Purpose & Method & Dataset\\
\noalign{\smallskip}\hline\noalign{\smallskip}

\cite{9} & 2014 & Present an approach to utilize this & LDA, Twitter-LDA \cite{80} & DBLP dataset\\
         & & valuable information source to suggest\\
          & & scientific articles\\

\cite{2}  & 2017 & A scientific paper recommendation approach &  LDA, Gibbs sampling   & ArnetMiner Dataset\\
     &   &   &      & DBLP dataset\\

\cite{3}  & 2017  & A Citation recommendation & LDA                           & ACL Anthology \\
     &       & present a topic model     &  Maximum A Posteriori (MAP)    & Network ,\\
     &       & combing with author link  &                                & DBLP \\
     &       & community\\

\cite{4}  & 	2013 & 	A personalized recommendation & LDA, EM algorithm & 	Digg articles\\
     &       &   system for Digg articles     & 	              &     (digg.com)\\

\cite{7} & 	2011 & 	A scientific articles recommendation & LDA, EM algorithm & 	CiteULike Dataset\\
     &       &   to users based on both content and\\
     &       &   other users ratings\\

\noalign{\smallskip}\hline
\end{tabular}
}

\end{table}

Also, some researchers, introduced a combined model based on traditional collaborative filtering and topic modeling and designed a novel algorithm to scientific articles recommendation for users from an online community, called CTR model. They considered LDA to initialize the CTR model, Infact they combined the matrix factorization and the LDA model, and is shown their approach better than the recommendations based on matrix factorization. For evaluation and test, Used a large dataset from a bibliography sharing service (CiteULike) \cite{7}. Table 1, shown some impressive work based on LDA for paper recommendation.\\

 \subsection{Topic model based on music and video recommendation}

Video and music recommendation has become an essential way for helping people explore the video world and discover the ones that may be of interest to them. Recently, analyze user interests and a good video or music recommendation in internet society is a big challenge. Hariri and et al. proposed a combined approach based on content and collaborative filtering methods from the sequence of songs listened to generate music recommendation. They applied a LDA model to reduce the dimensionality of the feature and obtain the hidden relationships between songs and tags. They collected 218,261 distinct songs from "Art of the Mix" website for evaluation their approach \cite{20}. Yan and et al, focused on the efficiency of users' information content on the online social network and provided a solution as a personalized video recommendation with considering users' cross-network social and content data. They applied a topic model based on LDA for each user, that user as document and user's hashtags as word, with considering user's information from Twitter \cite{12}. They derived Twitter user topic distributions and observed user-video interactions on YouTube, and presented a solution for user preference transfer:

\[{}^{min\ \ \ \ \ \ \ }_{v_{j.}W_{1.}W_{2.}}{\sum_{(i,j)\in \mathrm{\Omega }}{(r_{ij}-(\eta .u^{T_t}_iW_1+\left(1-\eta \right).u^{T_f}_iW_2)v^T_j)^2}} +\]
\[\theta \sum_j{F\left(v_j\right)+\lambda (||W_1{||}^2_F+||W_2{||}^2_F+\ \sum_j{||V_j{||}^2_F})}\]
\begin{equation} \label{GrindEQ__6_}
F\left(v_j\right)\triangleq {v_j\left(V^TL_j\right)+\left(V^TL_j\right)}^Tv^T_j-v_jL_{jj}v^T_j
\end{equation}

 $u^{T_t}_i$, Twitter user tweet topic distribution matrix $u^{T_t}$ = $\{$$u^{T_t}_i$ ;{\dots};$\ u^{T_t}_{|U|}$ $\}$,
$u^{T_f}_i,$ Twitter user social topic distribution matrix $u^{T_f}$= $\{$$u^{T_f}_i$ ;{\dots};$\ u^{T_f}_{|U|}$ $\}$,
 $W_1$and $W_2,\ $with observations of the overlapped user's Twitter and YouTube behaviors,
 $\mathrm{\Omega }$ as the collection of all the observed user-video pairs,
 $\eta $ is a trade-off parameter to balance the contribution of different types of user's behaviors on Twitter,
 where $v_j$ is the $j^{th}$row of $V$ , $L_j$ is the $j^{th}$column of \textit{L, }$L_{jj}$ is the entry located in the $j^{th}$column and $j^{th}$ row of \textit{L}.
 Based on, updated $v_j$ and $W_1$, $W_2$ iteratively until convergence or maximum iteration. The update rules are:
\[V_j\leftarrow V_j-\gamma \frac{d}{dW_2},   \ \ W_1\leftarrow W_1-\gamma \frac{d}{dW_1},    W_2\leftarrow W_2-\gamma \frac{d}{dW_2}, \]

where $\gamma$ denotes the learning rate.

 With the derived transfer matrices $W_1$, $W_2$ and video latent factor representations V, given a test user $u_i$ with his/her tweeting activity, friend collection, and the corresponding Twitter topical distributions $u^{T_t}_i$ , $u^{f_i}_i$ , we can estimate $u_i$  preferences on YouTube videos as:

 \begin{equation} \label{GrindEQ__7_}
R^{(1)}_{i,}=(\eta .u^{T_t}_iW_1\mathrm{+}\left(\mathrm{1-}\eta \right)\mathrm{.\ }u^{T_f}_iW_2)V^T
\end{equation}\\

For test and experiment, utilized their approach on YouTube-Twitter dataset
that include 9,253,729 tweeting behaviors and 1,097,982 video-related behavior
and showed combining auxiliary network information and utilizing a cross-network
Collaborative can lead to generating novel recommendations and increasing
satisfaction for users.

In addition, some researchers; proposed an approach based on Collaborative Filter (CF) Algorithm and utilized the application of session variety and temporal context. They applied a LDA model to temporal properties extraction of sessions and that considering sessions as documents and songs as words. For evaluation of this approach, they used \textit{Last.fm }dataset(log) that includes 19,150,868 entries from 992 users. Results showed that the approach with using temporal information can increase the accuracy of music recommendations \cite{14}.

In addition, other researchers used a dynamic framework based on four aspects of user's preference (collaborative aspect, content aspect, popularity aspect, and randomization aspect) for movie recommendation. The authors applied a linear combination model to generate the final recommendation list \cite{11}. Hu et al, proposed a novel topic modeling to audio retrieval, called GaussianLDA. In general, in this approach it was assumed that each audio document includes various latent topics and each topic considered as a Gaussian distribution. They prepared 1214 audio documents (length: between 0.82 s to 1 min), that each audio document is related with a category in different subject that includes: bell, river, rain, laugh, dog, gun and so on. Their results showed that the GaussianLDA model significantly outperform the standard LDA topic model\cite{16}. Table 2, shown some impressive work based on LDA for music and video recommendation.

% For tables use - t2
\begin{table}

\centering
\caption{Some impressive works LDA-based on Music and Video recommendation}
\label{tab:1}       % Give a unique label

\resizebox{12cm}{!} {
\begin{tabular}{lllll}
\hline\noalign{\smallskip}
Study & Year & Purpose & Method & Dataset\\
\noalign{\smallskip}\hline\noalign{\smallskip}

\cite{20} & 	2012 & 	A Music recommendation,        & LDA                      & Art of the Mix\\
     &       &  tracks and detects changes     & Collaborative filtering  & [www.artofthemix.org]\\
     &       &  in users's preferences \\

\cite{11} & 	2016 & 	A venue-aware music recommender &  	LDA, SVM & 	A music dataset:\\
     &       &   to identify suitable songs     &            & A music dataset:Concept-Labeled Music (TC1) \\
     &       &                                  &            & and Large Music (TC2)\\

\cite{12} & 	2016 & 	A video recommendation,         & 	LDA      & 	A Google+ dataset \\
     &       &  obtain users's rich social        &            &  (137,317 Google+ users)\\
     &       &  media behaviors\\

\cite{14} & 	2013 & 	A Music recommendation,         & LDA        & Last.fm\\ 	
     &       &  analysis of  user listening     &            & Collaborative filtering\\
     &       &  Behavior\\

\cite{15} & 	2010 & 	A Music recommendation,         &  LDA,       & 	A Zune Social music community\\
     &       &  characterizing user preferences &  Message-passing\\
     &       &  in social media content\\

\cite{16} & 	2014 & 	An audio retrieval,            & 	LDA,           &  An audio dataset\\
     &       &  Audio analysis of              &    Gaussian-LDA   &  in various categories\\
     &       &  multimedia information\\

\noalign{\smallskip}\hline
\end{tabular}
}

\end{table}

\subsection{Topic model based on location recommendation}
Recommendation systems based on location can suggest a set of places that users may be interested in, based on their history and behavior analysis. LDA can also be used for location recommendation. Kurashima et al, proposed a novel topic model for recommending new locations to visit, called Geo Topic Model. This model can predict user's interest and the user's spatial area based on features of visited locations. They used Tabelog-based (tablelog.com) and Flickr-based real-location log data for evaluation of their approach. They found that this model can discover latent topics related to art, great views, nature, atmosphere, and construction and other from logs of visited places \cite{24}. Table 3, shown some impressive work based on LDA for location recommendation.\\

% For tables use - t3
\begin{table}

\centering
\caption{Some impressive works LDA-based on location recommendation}
\label{tab:1}       % Give a unique label

\resizebox{12cm}{!} {

\centering
\begin{tabular}{lllll}
\hline\noalign{\smallskip}

Study & Year & Purpose & Method & Dataset\\
\noalign{\smallskip}\hline\noalign{\smallskip}

\cite{24} & 	2013 & 	A Location Recommendation, & 	LDA,         & 	A tabelog and  \\
     &       &  analysis of user's, Interest &  EM algorithm &   a Flick dataset\\
     &       &  and the user's spatial area  &               & [tabelog.com, Flickr.com]\\
     &       &  of activity\\

\cite{22} & 	2013 & 	Present a topic and               & 	LDA,           & 	A large real-world LBSN,\\
     &       &  location aware, Point-of-Interest & Gibbs sampling     &     Foursquare Dataset\\
     &       &  recommender system\\

\cite{26}  & 	2017 & 	A Spatial Item Recommendation, & LDA , Gibbs Sampling, & 	Foursquare Dataset,\\
     &       &  Study on patterns of           & a gradient descent learning  & A twitter dataset\\
     &       & users' behaviors                & algorithm\\

\noalign{\smallskip}\hline
\end{tabular}

}

\end{table}

Liu and et al, investigated the POI recommendation issue in LBSNs by mining textual information and proposed a 'Topic and Location-aware' probabilistic matrix factorization (TL-PMF) method for Point-of-Interest recommendation to discover personalized recommendations from favorite places  \cite{22}. The distribution over the observed ratings as well as the textual information is:

\begin{equation} \label{GrindEQ__9_}
p\left(R\mathrel{\left|\vphantom{R U,C,TL,{\sigma }^2}\right.\kern-\nulldelimiterspace}U,C,TL,{\sigma }^2\right)=\prod^M_{i=1}{\prod^N_{j=1}{[\mathcal{N}\left({\boldsymbol{r}}_{\boldsymbol{ij}}\right|f(U_i,C_{j,{TL}_{ij}}),{\boldsymbol{\sigma }}^{\boldsymbol{2}}\boldsymbol{)}]^{{\boldsymbol{I}}_{\boldsymbol{ij}}}}}
\end{equation}

Where, ${\boldsymbol{r}}_{\boldsymbol{ij}}$\textbf{ }be the rating of user\textbf{ }${\boldsymbol{u}}_{\boldsymbol{i}}$\textbf{ }for ${\boldsymbol{POI}}_{{\boldsymbol{c}}_{\boldsymbol{j}}}{\boldsymbol{C}}_{\boldsymbol{j}}$,

${\boldsymbol{U}}_{\boldsymbol{i}}$and ${\boldsymbol{C}}_{\boldsymbol{i}}$\textbf{ }are the user and \textbf{POI} latent feature space vector respectively,

\textbf{ }$\mathcal{N}\left(\boldsymbol{\mathrm{\textrm{-}}}\right|,{\boldsymbol{\sigma }}^{\boldsymbol{2}}\boldsymbol{)}$ is a Gaussian distribution with mean  and variance ${\boldsymbol{\sigma }}^{\boldsymbol{2}}$,

${\boldsymbol{I}}_{\boldsymbol{ij}}$\textbf{ }is the indicator function,

Function $f(U_i,C_{j,{TL}_{ij}})$ is to approximate the rating of user $u_i$ for ${POI}_{cj}$.

they analyzed the topic characteristics from POIs across various geographical areas.
The experiments were conducted on a large real-world LBSN dataset; they analyzed the topic characteristics from POIs across various geographical areas.

\subsection{Topic model based on friend recommendation}
Friend recommendation is a popular method to help users to make new friends and discover interesting information. Friend recommendation is a relative challenging issue contrasted with group or item recommendations in online social networks\cite{69,79,80}. To address this challenge issue, a recent work in [79] proposed a friend recommendation based on LDA, which contains two stages: first step, they applied tag-user information to produce a possible friend list and then they created a topic model to demonstrate the relationship between user's friend making behaviour and  image features. They applied experiments on the Flickr as a standard dataset and showed that their recommends friends more quickly than traditional methods.

\subsection{Topic model Based on travel and tour recommendation}

Definitely, recommendation systems can have a significant impact to build a smart travel recommendation. Many different techniques have recently been developed to support travel recommendation based on different kinds of data. For example, in \cite{49}, the authors proposed a novel generative probabilistic model named socoLDA with heterogeneous social influence to better capture users's travel interests. They introduced the framework of travel-package recommendation named socoTraveler, which applys socoLDA to show a user's travel interests in topic space and find similar users to produce recommendations with user-based collaborative filtering. In addition, in \cite{51}, the authors proposed a framework to suggest top-k tours with highest marks for a user by using the photos shared by other users in an online social network.  Table 4, shown some impressive work based on LDA for travel and tour recommendation.

% For tables use - t4
\begin{table}

\centering
\caption{Some impressive works LDA-based on Travel and Tour recommendation}
\label{tab:1}       % Give a unique label
\resizebox{12cm}{!} {

\centering
\begin{tabular}{lllll}
\hline\noalign{\smallskip}
Study & Year & Purpose & Method & Dataset\\
\noalign{\smallskip}\hline\noalign{\smallskip}

\cite{49} & 	2016 & 	A travel-package                & LDA,           & A real travel dataset\\
     &       &  recommendations with            & Gibbs sampling & (from a China tourism\\
     &       &  considering social influence    &                &  company)\\

\cite{50} & 	2017 & 	A tourism recommendation        & LDA,           & 	A twitter dataset\\
     &       &         on twitter users         & Gibbs sampling\\

\cite{51}  & 	2017 & 	A Tour recommendations          & LDA,           &  A Flickr dataset\\
     &       &                                  & Gibbs sampling\\

\cite{52}  & 	2015 & 	Personalized trip &  LDA,                    & 	Yelp and\\
     &       &  recommendation    &  collaborative filtering & Foursquare dataset\\

\cite{27}  & 	2017 & 	A social sequential  & 	LDA,  & 	A twitter dataset \cite{81}\\
     &       &  tour based POI       &  Gibbs sampling\\
     &       &  (point of interest)  &   \\

\cite{82}  & 	2016 & 	Trip recommendation & 	LDA, Yelp and\\
     &       &  based on points of  &  Depth-first search (DFS)  &  Foursquare dataset\\
     &       &  interest (POIs)\\

\noalign{\smallskip}\hline
\end{tabular}

}

\end{table}

\subsection{Topic model based on app recommendation}
Currently, a wide range of recommendation approaches have been proposed and applied to recommend mobile apps. For example, In \cite{55}, proposed a novel probabilistic model, named Goal-oriented Exploratory Model (GEM), to combine the identification of exploratory behavior and mobile app recommendations into a unified framework. The authors, employed the idea of LDA to design a topic model to cluster items into goals and identify the personal distribution of goals for each user and developed an effective and efficient algorithm, which integrates Expectation-Maximization (EM) algorithm with collapsed Gibbs sampling for model learning. They collected a mobile app dataset from Qihoo 360 Mobile Assistant, an open mobile app platform in China for Android users. Lin et al, investigated the cold-start issue with using the social information for App recommendation in Twitter and used a LDA model to discovering latent group from "Twitter personalities" to recommendations discovery. This approach is based on a simple "averaging" method where the probability of how likely the target user will like the app is the expectation of how the Twitter-followers like the app. Given a set of Twitter-followers $T$, the probability that user $u$ likes app a is defined as follows:

\begin{equation} \label{GrindEQ__10_}
p\left(+\mathrel{\left|\vphantom{+ a,u}\right.\kern-\nulldelimiterspace}a,u\right)=\sum_{t\in T(a)}{p(+,t|a,u)}=\sum_{t\in T\left(a\right)}{p\left(+\mathrel{\left|\vphantom{+ t,u}\right.\kern-\nulldelimiterspace}t,u\right)}P(t|a)
\end{equation}

where \textit{T(a)} is the set of possible Twitter-followers following app
\textit{a}, in which assume that:

\textbf{(i)} Twitter-followers are examined once at a time to make a decision
about whether an app is liked or disliked.

\textbf{ (ii)} when the Twitter follower is known, the judgement does not depend
on the app any more, i.e.,

\textbf{(iii)} the fact that given a user and an app, there is no judgement
involved, i.e.,

\textbf{(iv)} the fact that an app has a given Twitter follower is independent
from the user, i.e.,

 Equation \eqref{GrindEQ__11_} is then reduced to the estimation of two quantities:

 1. The probability that user u likes app a given that app a has Twitter-follower t, i.e., $p(+|t,\ u)$, and

 2. The probability of considering Twitter-follower t given app a, i.e., $p(t|a)$

$p(+|t,\ u)$ is straightforward to estimate as it can be rewritten as:
\begin{equation} \label{GrindEQ__11_}
p\left(+\mathrel{\left|\vphantom{+ t,u}\right.\kern-\nulldelimiterspace}t,u\right)=\frac{p(+,t|u)}{p\left(+,t\mathrel{\left|\vphantom{+,t u}\right.\kern-\nulldelimiterspace}u\right)+p(-,t|u)},
\end{equation}
where $p(+,\ t|u)$ and $p(-,\ t|u)$ are derived from LDA, which is the probability that Twitter-follower t occurs in an app that is liked (or disliked) by user u.

For test and experiment, they considered Apple's iTunes App Store and Twitter as dataset. Experimental results show, their approach significantly better than other state-of-the-art recommendation techniques  \cite{83}.

In  \cite{56}, the authors proposed an allocation-based probabilistic mechanism that considers multiple user-app factors to help users with app recommendations. This framework that can capture geographical influences on usage behaviors and effectively model user mobility patterns, which in turn affect app usage patterns. They used Gibbs sampling to approximately estimate and infer the parameters of LDA. The authors  measure the similarity of two location blocks by the similarity of app usage pattern, which is calculated by Pearson's correlation similarity \cite{84}:

\begin{equation} \label{GrindEQ__12_}
Sim(l_x,\ l_y)=\frac{\sum_{{\alpha }_j\in A}{(I_{aj,l_x}-\overline{I_{l_x}})(I_{aj,l_y}-\overline{I_{l_y}})}}{\sqrt{\sum_{{\alpha }_j\in A}{(I_{aj,l_x}-\overline{I_{l_x}})^2(I_{aj,l_y}-\overline{I_{l_y}})^2}}}
\end{equation}
Specifically,

\noindent  defined $c({\mu }_i,{\alpha }_j,l_x)$=1, if user ${\mu }_i$has launched app ${\alpha }_j$at location block $l_x$.

\noindent Otherwise, $c({\mu }_i,{\alpha }_j,l_x)$=0. Therefore, the number of users who have launched ${\alpha }_j$ at location block $l_x$ is  $I_{aj,l_x}=\sum_{{\mu }_i\in U}{c({\mu }_i,{\alpha }_j,l_x)}$\textbf{.}

\noindent where $\overline{I_{l_x}}$ denotes average mobile app influences at location block $l_x$.

\noindent To determine whether a location block $l_z$ belongs to a location $L$,

\noindent

\noindent They define the app usage pattern coefficient as $r_{l_x}$for geographical block $l_x$. The initial collaborative filter coefficient is the average mobile app influence. To decide whether location block $l_z$ belongs to geographical region $L$. Also, they calculate its collaborative filter coefficient as follows :

\begin{equation} r_{l_{z}}= \frac{1}{s_{L}}\sum\limits_{l_{x}\in L}sim(l_{x}, l_{z})^\ast r_{l_{x}} \end{equation}

\noindent where ${\boldsymbol{s}}_{\boldsymbol{L}}$ is the number of location blocks in geographical region $L$. If the value of $r_{l_z}$\textbf{$\boldsymbol{>}$}$\ r_{th}$, where $r_{th}$\textbf{ }is a predefined threshold value, location block $l_z$will be included in $L$.

% For tables use - t5
\begin{table}

\centering
\caption{Some recent and impressive works LDA-based on app recommendation}
\label{tab:1}       % Give a unique label

\resizebox{12cm}{!} {
\centering
\begin{tabular}{lllll}
\hline\noalign{\smallskip}
Study & Year & Purpose & Method & Dataset\\
\noalign{\smallskip}\hline\noalign{\smallskip}

\cite{53}  & 	2017 & 	A mobile App recommendation & 	LDA                & 	A benchmark dataset based upon\\
     &       &                              &  stochastic gradient &    Apple's iTunes App Store\\
     &       &                              &  descent (SGD)\\

\cite{54}  & 	2017 & 	A mobile App recommendation &  LDA,               & iphone-iPad ,\\
     &       &  with considering users and  & stochastic gradient & iphone-iPad-iMac Dataset\\
     &       &  Apps data on multiple      & descent (SGD),      &  (manual obtained)\\
     &       &                              &  Collaborative Topic modeling (C. Wang \& Blei, 2011)\\

\cite{55}  & 	2017 & 	A mobile app recommendation & LDA,               & A  mobile app dataset\\
     &       & with exploratory behavior    & EM Collapsed Gibbs &(Qihoo 360 Mobile\\
     &       & from big data                & sampling algorithm & Assistant in China)\\

\cite{56}  & 	2017 & 	A mobile application recommendation    & LDA,           & A app mobile dataset\\
     &       & with considering geographical location  & Gibbs Sampling & by a chinese app\\
     &       &                                         &                & (AntTest)\\

\cite{57}  & 	2016 & 	A semantic recommendation              & LDA,           & A app dataset\\
     &       &  to evaluate mobile applications        &Gibbs Sampling  & of iOS mobile \\
     &       &  with behavior Analysis  	           &                & apps\\

\cite{83}  & 	2013 & 	Addressing the Cold-Start              & LDA                      & A dataset from Apple's iTunes\\
     &       &  problem, A app Recommendation          & Gibbs Sampling           & App Store\\
     &       &  with considering Twitter-followers     & collaborative filtering  & and Twitter\\

\noalign{\smallskip}\hline
\end{tabular}

}

\end{table}

\section{Experiment: Semantic Discovery and researchers behavior analysis}
\subsection{ Dataset}

We extracted ISWC and WWW conferences publications from DBLP website by only considering conferences for which data was available for years 2013-2017. In total, It should be noted that in these experiments, we considered abstracts and titles from each article. In this paper, we used MALLET (http://mallet.cs.umass.edu/) to implement the inference and obtain the topic models. In addition, our full dataset is available at https://github.com/JeloH/Dataset\_DBLP. The most important goal of this experiment is discover the trends of the topics and find relationship between LDA topics and paper features and generate trust tags.

\subsection{Parameter Settings}
 In this paper, all experiments were carried out on a machine running Windows 7 with CoreI3 and 4 GB memory. We learn a LDA model with 100 topics; $\alpha=0.01$, $\beta = 0.01$ and using Gibbs sampling as a parameter estimation. Related words for a topic are quite intuitive and comprehensive in the sense of  supplying a semantic short of a specific research field.

 \subsection{Semantic analysis and generate tags}
In this section, we provide the results and discovered topics of 100-topics for ISWC and WWW conferences.

%%%%%%%%%%%%%%%%%%%%%%%%%%%%%%%%%%%%%%--Table: ISWC and WWW --%%%%%%%%%%%%%%%%%%%%%%%%%%%%%%%%%%%%%%%%%%%%%%%%%%%
\begin{table}[]
\centering
\caption{This result discovered topics of 100-topics from ISWC and WWW conferences, each topic is shown with the top 20 words and identified 'fields of research' that are covered in this conference. Topics that have two concepts are in blue and red colors.}
\label{my-label}

\resizebox{12cm}{!} {
\begin{tabular}{|l|l|l|l|c|l|c|l|}
\hline
\multicolumn{4}{|c|}{ISWC}                                                                                                                                                                                                                                                                                                                                                                                                                                                                                                        & \multicolumn{4}{c|}{WWW}                                                                                                                                                                                                                                                                                                                                                                                                                                                                                                                                                                                                                                             \\ \hline
\multicolumn{2}{|l|}{Semantic Web}                                                                                                                                                                                                         & \multicolumn{2}{l|}{Ontology modularity}                                                                                                                                                                                                                                                    & \multicolumn{2}{c|}{Web content  analysis }                                                                                                                                                                                                                                                                   & \multicolumn{2}{c|}{\begin{tabular}[c]{@{}c@{}}{\color{blue}Social search and mining}, {\color{red}Crowdsourcing}\end{tabular}}                                                                                                                                                                                                                            \\ \hline
\multicolumn{2}{|c|}{Topic 14}                                                                                                                                                                                                              & \multicolumn{2}{c|}{Topic 28}                                                                                                                                                                                                                                                             & \multicolumn{2}{c|}{Topic 87}                                                                                                                                                                                                                                                                                      & \multicolumn{2}{c|}{Topic 61}                                                                                                                                                                                                                                                                                                                    \\ \hline

\begin{tabular}[c]{@{}l@{}}data\\semantic\\web\\based\\query\\ontologies\\present\\rdf\\evaluation\\system\end{tabular} &
 \begin{tabular}[c]{@{}l@{}}open\\graph\\provide\\search\\analysis\\processing\\users\\domain\\semantics\\number\end{tabular} &

 \begin{tabular}[c]{@{}l@{}}linked\\approach\\ontology\\entity\\entities\\datasets\\owl\\systems\\performance\\language\end{tabular} &
 \begin{tabular}[c]{@{}l@{}}framework\\set\\approaches\\techniques\\methods\\method\\technologies\\quality\\existing\\efficient\end{tabular} &

 \multicolumn{1}{r|}{\begin{tabular}[c]{@{}r@{}}data\\users\\web\\based\\information\\paper\\problem\\study\\models\\analysis\end{tabular}} & \begin{tabular}[c]{@{}l@{}}set\\systems\\existing\\user\\task\\important\\structure\\approaches\\topics\\effectiveness\end{tabular} &

  \multicolumn{1}{r|}{\begin{tabular}[c]{@{}r@{}}{\color{red}model}\\{\color{blue}search}\\show \\{\color{blue}network} \\propose \\{\color{blue}online} \\query \\applications \\{\color{blue}queries} \\tasks\end{tabular}} & \begin{tabular}[c]{@{}l@{}}{\color{red}patterns}\\{\color{red}understanding} \\algorithms \\prediction\\ specific \\{\color{red}language} \\{\color{blue}mining} \\domains\\{\color{blue}crowdsourcing} \\properties\end{tabular} \\ \hline

%%%%%%%%%%%%%%%%%%%%%%%%%%%%%%%%%%%%%%%%%%%%%%%%%%%%%%%%%%%%%%%%%%%

\multicolumn{2}{|l|}{{\color{blue}Mapping}, {\color{red} Query language}}                                                                                                                                                                                                         & \multicolumn{2}{l|}{{\color{blue} Cloud computing}, {\color{red}Data stream} }                                                                                                                                                                                                                                                    & \multicolumn{2}{c|}{online social networks}                                                                                                                                                                                                                                                                   & \multicolumn{2}{c|}{\begin{tabular}[c]{@{}c@{}}question answering and social media \end{tabular}}                                                                                                                                                                                                                            \\ \hline
\multicolumn{2}{|c|}{Topic 57}                                                                                                                                                                                                              & \multicolumn{2}{c|}{Topic 4}                                                                                                                                                                                                                                                             & \multicolumn{2}{c|}{Topic 18}                                                                                                                                                                                                                                                                                      & \multicolumn{2}{c|}{Topic 7}                                                                                                                                                                                                                                                                                                                    \\ \hline

\begin{tabular}[c]{@{}l@{}}{\color{blue}knowledge}\\{\color{red}sparql} \\{\color{red}queries} \\show \\{\color{blue}time} \\{\color{blue}propose} \\real \\reasoning \\{\color{blue}state} \\{\color{blue}mappings} \end{tabular} &
 \begin{tabular}[c]{@{}l@{}}world\\{\color{red}schema}\\existing \\{\color{red}scale} \\experiments \\{\color{red}dl} \\application \\multiple \\{\color{red}benchmark} \\{\color{blue}tasks}\end{tabular} &

 \begin{tabular}[c]{@{}l@{}}paper\\{\color{blue}information}\\results \\{\color{blue}large} \\{\color{red}rdf} \\{\color{red}model} \\{\color{blue}user} \\{\color{blue}applications} \\{\color{red}context} \\learning\end{tabular} &
 \begin{tabular}[c]{@{}l@{}}problem\\task\\algorithms\\{\color{red}content}\\{\color{red}languages} \\wikipedia \\{\color{blue}cloud} \\order \\{\color{red}stream} \\{\color{blue}algorithm}\end{tabular} &

 \multicolumn{1}{r|}{\begin{tabular}[c]{@{}r@{}}time\\results\\approach\\networks\\algorithm\\work\\scale\\graph\\methods\\method\end{tabular}} & \begin{tabular}[c]{@{}l@{}}people\\content\\entities\\develop\\online\\significantly\\key\\optimal\\investigate\\previous\end{tabular} &

  \multicolumn{1}{r|}{\begin{tabular}[c]{@{}r@{}}social\\framework\\features\\demonstrate\\learning\\question\\ranking\\services\\modeling\\accuracy\end{tabular}} & \begin{tabular}[c]{@{}l@{}}order\\common\\potential\\size\\ad\\activity\\generated\\proposed\\browser\\relevant\end{tabular} \\ \hline

 %%%%%%%%%%%%%%%%%%%%%%%%%%%%%%%%%%%%%%%%%%%%%%%%%%%%%%%%%%%%%%%%%%%

\multicolumn{2}{|l|}{ Question Answering System }                                                                                                                                                                                                         & \multicolumn{2}{l|}{Concept representation}                                                                                                                                                                                                                                                    & \multicolumn{2}{c|}{Recommendation  systems }                                                                                                                                                                                                                                                                   & \multicolumn{2}{c|}{\begin{tabular}[c]{@{}c@{}}Community detection \end{tabular}}                                                                                                                                                                                                                            \\ \hline
\multicolumn{2}{|c|}{Topic 59}                                                                                                                                                                                                              & \multicolumn{2}{c|}{Topic 25}                                                                                                                                                                                                                                                             & \multicolumn{2}{c|}{Topic 19}                                                                                                                                                                                                                                                                                      & \multicolumn{2}{c|}{Topic 29}                                                                                                                                                                                                                                                                                                                    \\ \hline

\begin{tabular}[c]{@{}l@{}}sources\\ontology\\evaluate\\terms\\models\\relations\\answering\\describe\\current\\improve\end{tabular} &
 \begin{tabular}[c]{@{}l@{}}relevant\\related\\main\\years\\impact\\find\\statistical\\corpus\\means\\enables\end{tabular} &

 \begin{tabular}[c]{@{}l@{}}tools\\representation\\test\\structure\\quality\\investigate\\addition\\detection\\feature\\part\end{tabular} &
 \begin{tabular}[c]{@{}l@{}}values\\functions\\annotations\\concept\\chinese\\alignments\\multi\\media\\outperforms\\resulting\end{tabular} &

 \multicolumn{1}{r|}{\begin{tabular}[c]{@{}r@{}}user\\real\\world\\recommendation\\datasets\\art\\state\\experiments\\level\\items\end{tabular}} & \begin{tabular}[c]{@{}l@{}}semantic\\influence\\find\\product\\recent\\human\\wikipedia\\machine\\outperforms\\platform\end{tabular} &

  \multicolumn{1}{r|}{\begin{tabular}[c]{@{}r@{}}present\\number\\context\\questions\\find\\multiple\\community\\clustering\\group\\evaluation\end{tabular}} & \begin{tabular}[c]{@{}l@{}}experimental\\relevance\\similar\\proposed\\describe\\highly\\rating\\factorization\\space\\detect\end{tabular} \\ \hline

  %%%%%%%%%%%%%%%%%%%%%%%%%%%%%%%%%%%%%%%%%%%%%%%%%%%%%%%%%%%%%%%%%%%

\multicolumn{2}{|l|}{documents analysis }                                                                                                                                                                                                         & \multicolumn{2}{l|}{Linked Data, semantically query }                                                                                                                                                                                                                                                    & \multicolumn{2}{c|}{Social network and security}                                                                                                                                                                                                                                                                   & \multicolumn{2}{c|}{\begin{tabular}[c]{@{}c@{}}{\color{blue}User interaction}, {\color{red}Query language } \end{tabular}}                                                                                                                                                                                                                            \\ \hline
\multicolumn{2}{|c|}{Topic 20}                                                                                                                                                                                                              & \multicolumn{2}{c|}{Topic 58}                                                                                                                                                                                                                                                             & \multicolumn{2}{c|}{Topic 92}                                                                                                                                                                                                                                                                                      & \multicolumn{2}{c|}{Topic 69}                                                                                                                                                                                                                                                                                                                    \\ \hline

\begin{tabular}[c]{@{}l@{}}documents\\version\\highly\\scientific\\networks\\web\\health\\bases\\role\\constructs\end{tabular} &
 \begin{tabular}[c]{@{}l@{}}correctness\\multilingual\\path\\problems\\past\\structure\\performed\\kg\\tips\\commercial\end{tabular} &

 \begin{tabular}[c]{@{}l@{}}dataset\\source\\graphs\\access\\high\\dbpedia\\rules\\alignment\\human\\triples\end{tabular} &
 \begin{tabular}[c]{@{}l@{}}cost\\project\\engine\\complex\\generation\\empirical\\relationships\\named\\computing\\space\end{tabular} &

 \multicolumn{1}{r|}{\begin{tabular}[c]{@{}r@{}}content\\twitter\\groups\\attacks\\google\\link\\page\\training\\tracking\\matrix\end{tabular}} & \begin{tabular}[c]{@{}l@{}}click\\path\\results\\role\\attack\\constraints\\leads\\urls\\accounts\\presents\end{tabular} &

  \multicolumn{1}{r|}{\begin{tabular}[c]{@{}r@{}}{\color{blue}user}\\{\color{blue}pages}\\address \\{\color{blue}news} \\factors \\{\color{blue}sources} \\{\color{red}SPARQL} \\{\color{blue}communication} \\facebook \\{\color{red}business}\end{tabular}} & \begin{tabular}[c]{@{}l@{}}evaluate \\random \\{\color{red}agents} \\extremely \\measurement \\exploiting \\high \\{\color{blue}clicks} \\{\color{red}hierarchical} \\{\color{red}computational}\end{tabular} \\ \hline

\multicolumn{2}{|l|}{semantic network and  natural language}                                                                                                                                                                                                         & \multicolumn{2}{l|}{RDFa for Educational websites}                                                                                                                                                                                                                                                    & \multicolumn{2}{c|}{Mobile Networks}                                                                                                                                                                                                                                                                   & \multicolumn{2}{c|}{\begin{tabular}[c]{@{}c@{}}User behavior\end{tabular}}                                                                                                                                                                                                                            \\ \hline
\multicolumn{2}{|c|}{Topic 52}                                                                                                                                                                                                              & \multicolumn{2}{c|}{Topic 66}                                                                                                                                                                                                                                                             & \multicolumn{2}{c|}{Topic 14}                                                                                                                                                                                                                                                                                      & \multicolumn{2}{c|}{Topic 24}                                                                                                                                                                                                                                                                                                                    \\ \hline

\begin{tabular}[c]{@{}l@{}}selection\\technical\\network\\algebra\\comprehensive\\clinical\\natural\\logs\\extensions\\contrast\end{tabular} &
 \begin{tabular}[c]{@{}l@{}}names\\tabular\\deployed\\strategies\\represent\\limitation\\networks\\citybench\\generalized\\nodes\end{tabular} &

 \begin{tabular}[c]{@{}l@{}}usability\\benchmarking\\suite\\valuable\\shown\\apply\\exploring\\feasible\\lessons\\tackle\end{tabular} &
 \begin{tabular}[c]{@{}l@{}}concepts\\educational\\systematic\\supervised\\bases\\rdfa\\mechanism\\fly\\taxonomical\\annotating\end{tabular} &

 \multicolumn{1}{r|}{\begin{tabular}[c]{@{}r@{}}large\\mobile\\provide\\system\\research\\evaluate\\improve\\effective\\studies\\make\end{tabular}} & \begin{tabular}[c]{@{}l@{}}low\\interest\\focus\\analyze\\global\\networks\\accurate\\achieve\\hand\\distributed\end{tabular} &

  \multicolumn{1}{r|}{\begin{tabular}[c]{@{}r@{}}behavior\\general\\location\\dynamics\\articles\\evidence\\words\\evolution\\activities\\ratings\\complete\end{tabular}} & \begin{tabular}[c]{@{}l@{}}intent\\rate\\video\\class\\gender\\consumption\\dataset\\step\\experts\end{tabular} \\

   \hline

\end{tabular}
}

\end{table}

According to Table 6 , the following observations can be made:

In ISWC conference, Topic 25 sounds considerably more generic and is consistent with 'Concept Representation' in general, and marked by representation, structure, investigate, detection, values, and concept. Also, we can see that from 20 generated words in Topic 20, some words are very related to each other in means such as documents, scientific, networks, web, health, kg and we found that this topic covers papers that propose models  in 'Documents Analysis in health research'.\\

In WWW conference, Topic 7, this is our question; the word of 'ad'!  Is for 'Advertisement' or 'Ad Hoc Network'? As we can see, the word 'ad' can be related to 'Advertisement' or also to 'ad hoc network'. To answer this question, it is very easy to see that topic 7 reveals social, modeling, ranking, question, ad, browser. If only we consider the words 'browser', news, social, we can predict that this topic can be related to 'Advertisement' and this topic covers papers that propose methods in 'Question Answering and Social Media'.

\section{Discussion, Open Issues and Future Directions}
In this study, we focused on the LDA approaches to recommendation systems and given the importance of research, we have studied recent impressive articles on this subject and presented a taxonomy of recommendation systems based on LDA of the recent research. we evaluated ISWC and WWW conferences articles from DBLP website and used the Gibbs sampling algorithm as an evaluation parameter. We succeeded in discovering the relationship between LDA topics and paper features and also obtained the researchers' interest in research field. According to our studies, some issues require further research, which can be very effective and attractive for the future.

\subsection{Topic modeling methods and traditional methods in recommendation systems}
There are differences between recommendation systems based on LDA and traditional collaborative filtering (CF), we discus about the issue of 'Cold-start', 'latent user interest' and Sparsity in the field of recommender systems. It should be noted that to overcome the major weaknesses of CF-based recommendation systems, many models have been proposed, such as \cite{7}.

\textit{Recommendation systems based on LDA in cold-start}, the cold start problem occurs when a new item or user has just
    logged into that system; it is difficult to find similar ones because there is not enough information. LDA can be effective and useful to deal with cold-start in recommendation systems. There are approaches based on LDA to deal with this problem that some of the methods are combined with CF methods, for example, Lin et al, investigated the cold-start issue with use the social information for app recommendation in Twitter and used a LDA model to discovering latent group from 'Twitter personalities' to recommendations discovery and shown that their approach overcomes the difficulty of cold-start app recommendation. Also Some researchers investigated the cold-start problem in tag recommendation, for example; In \cite{20} , presented a system recommendation based on LDA for Cold-Start in Music Recommendation and showed that their approach can be useful in handling the cold start problem where a new song hasn't occurred in the training data. Other researchers also analyze the cold problem in the Video recommendation, for example; In \cite{12} introduced a unified YouTube video recommendation solution via cross-network collaboration and LDA to address the typical cold-start and data sparsity problems in recommender systems and show that this approach can be effective in terms of precision and improving the diversity of the recommended videos.\\

\textit{ Recommendation Systems based on LDA for 'Sparsity problem'}, the data sparsity challenge happens when the ratio is too small to supply enough information for effective predictions in CF systems, and hence the access matrix is very sparse. Recommendation systems based on LDA with ratings data can provide significant advantage and addition can be useful for exploratory data analysis and dimensionality reduction in huge content-text. This dimensionality reduction can also help to alleviate the sparseness problem which is inherent to many traditional collaborative-filtering systems. There are approaches based on LDA to deal with this Scarcity problem such as \cite{92}\\

\textit{Recommendation Systems based on LDA for 'user latent interests'}, latent interest refers to long-term interest in a specific topic, in fact; A latent interest can be viewed as one specific characteristic of the users and the items who have this latent interest will prefer the items with this characteristic. Finding a community with a latent interest in another could help in recommending interesting new communities for a user. However, for CF systems, it is hard to identify the user latent interests, since the only information available is the user interaction information with the system. While topic models can be utilized to simulate the user latent interests and showed the way of extracting these interests from the latent Dirichlet allocation (LDA) model by the Gibbs sampling method in our experiment. We should note that, in terms of items, a latent interest can be viewed as one specific characteristic of the items and the users who have this latent interest will prefer the items with this characteristic. Fortunately, as a simulation tool, the topic model (e.g., LDA) can be utilized to learn the meaning, significance, characteristics and attributes of items in a data-driven, i.e., from given rating records, possibly without further content or prior knowledge of these items\cite{118}.

\section{Conclusion} \label{sec:1}
In this paper, we presented a taxonomy of recommendation systems and applications based on LDA of the recent research, including app, travel, friend, location, scientific paper, and music recommendation. Furthermore, we applied LDA algorithm and Gibbs sampling on ISWC and WWW conference's publications from 2013-2017. Generally, recommendation systems can be an impressive interface between online users and websites in the Internet communities. Our study suggest that NLP methods based on LDA can discover hidden aspects to better understanding of the behaviors of the people to build smart recommendation systems in online communities.

\end{document}